\documentclass[
preprint,
 amsmath,amssymb,
 aps,
pre,
floatfix,
]{revtex4-2}

\usepackage{graphicx}
\usepackage{dcolumn}
\usepackage{bm}

\begin{document}

\preprint{APS/123-QED}

\title{ Antipolar Cell--cell Adhesion-causing Collective Motility Disorder}

\author{Katsuyoshi Matsushita}
 \email{kmatsu3@hiroshima-u.ac.jp}
\author{Koichi Fujimoto}%
 \email{kfjmt@hiroshima-u.ac.jp}
\affiliation{%
 Graduate School of Integral Life Science, Hiroshima University.
}%


\author{Mami Matsumoto and Kazunobu Sawamoto}
\affiliation{
Graduate School of Medical Sciences, Nagoya City University. \\
}%


\date{\today}

\begin{abstract}
In this study, we aim to theoretically investigate antipolar cell-cell adhesion, in which adhesion sites are located on the opposite side of the leading edge of migrating cells, as a candidate for irregularly polarized adhesion that induces disorder in collective cell migration. We employ the cellular Potts model to simulate the effects of antipolar adhesion on collective migration driven by cell motility. Antipolar adhesion induces a collective motility disorder, which exhibits a disordered configuration in the motility direction, even when collective motion occurs in the absence of adhesion. Consequently, antipolar adhesion inhibits collective migration.  The effect is in contrast to that of polar adhesion, which accelerates the directional intercellular order of cell motility. At a specific motility strength, a depinning transition emerges from a collective motility disorder to a collective motion. The collective motility disorder can be physically explained by the cooperative effect between antipolar adhesion and motility persistence within the mean-field approximation.
\end{abstract}

\maketitle


\section{\label{sec:introduction}Introduction}
Eukaryotic cells collectively migrate to their functional locations during developmental and physiological processes \cite{Weijer:2009}. In collective migration, the direction of cell motility, which appears as intracellular cytoskeletal orientation, chemical gradient, or organelle alignment on the leading-rear cell axis \cite{Mogilner:2012}, often exhibits an ordered configuration among cells \cite{Lois:1994, Haga:2005, Maeda:2008, Kabla:2012, Bertrand:2024}. The direction arises from intercellular interactions that inform the direction of motility through chemical signals \cite{Rappel:1999, McDonald:2003, Merks:2005, Merks:2008a, Merks:2008b, Palm:2013, Camley:2016, Varennes:2017}, mechanical force \cite{Trepat:2009, Reffey:2014, Kopf:2013, Sato:2015a, Kopf:2015, Yabunaka:2017b}, and cell-cell behavior during collisions \cite{Schneyder:2017, Matsushita:2019, Hiraiwa:2020, Hayakawa:2020, Khataee:2021} under the guidance of various extracellular cues \cite{Belvindrah:2007, Khalila:2010, Angelini:2010, Sawada:2013, Safran:2013, Ajioka:2015, Novikova:2017, Leoni:2017, Fujioka:2017, Kaneko:2018, Pinto:2022, Wang:2025}. Irregularity in these interactions results in disordered cell motility and poses serious obstacles to developmental and physiological processes. Theoretically, investigating the irregular interaction effects by constructing models from a physical perspective \cite{Camley:2017, Alert:2020} potentially provides insights into the mechanisms underlying the obstacles to these processes. Understanding irregular interactions that cause disordered cell motility can enable broad applications to a comprehensive understanding of phenomena, not only in biophysics and active matter physics \cite{Marchetti:2013, Hakim:2017} but also to developments in cell-based bioengineering and regenerative medicine \cite{Friedl:2009}.

\begin{figure}[t]
\begin{center}
\includegraphics[width=1.0\linewidth]{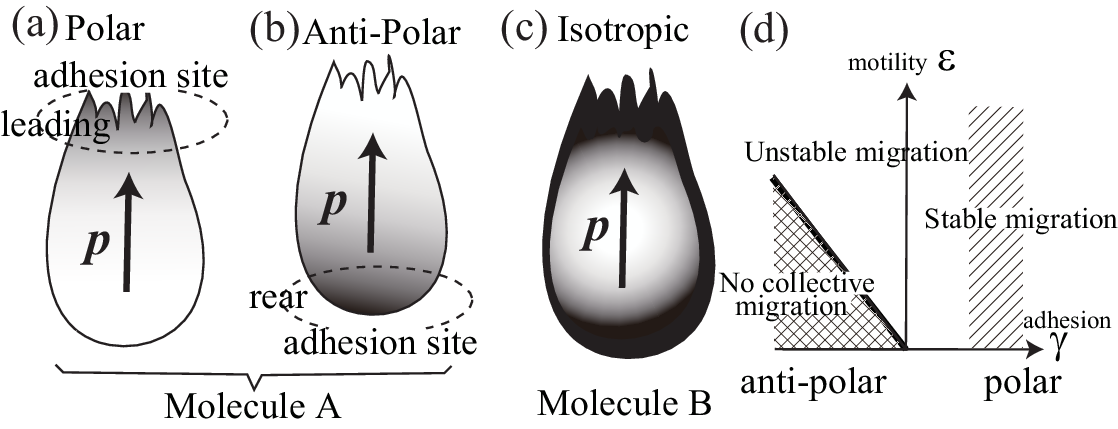}
\caption{ Schematic sketches of (a) polar adhesion, (b) antipolar adhesion, and (c) isotropic adhesion. In these panels, arrows with $\bm p$ represent the motile directions of cells, and the grey-colored regions represent the concentration of adhesion molecules for the corresponding cell-cell adhesion. The dark gray color indicates a high concentration of adhesion molecules. In the present study,  heterophilic adhesion is considered between cells with (a) and (c) or (b) and (c). The combination of adhesion is determined by the sign of adhesion strength $\gamma$. (d) Schematic phase diagram of multicellular behavior in the $\gamma$-$\varepsilon$ plane, where $\varepsilon$ represents the magnitude of cell motility. Stable collective migration and collective motility disorder appear in slashed and cross-hatched regions, respectively.   
}\label{fig:concept}
\end{center}
\end{figure}
One of the significant interactions affecting cells is mechanical contact through cell-cell adhesion \cite{Najem:2016, Pascalis:2017, Fujiike:2018, Fujimori:2024}. In particular, the non-uniform concentration of adhesion molecules on the cell surface induces various cell-sorting patterns \cite{Glazier:1993, Kafer:2007, Grazier:2008, Graizer:2011, Nakajima:2011, Takeichi:2014, Hirashima:2017, Belousov:2024, Braat:2024}, tissue elongation \cite{Zajac:2003, Belmonte:2016, Pfister:2016}, and migration phenomena \cite{Coates:2001, Matsushita:2018, Fujimori:2019, Noureen:2025}. An example of non-uniform concentration is the polarization in the leading edge direction $\bm p$, which is shown in Fig.~\ref{fig:concept}(a). Polarization in the concentration of adhesion molecules, henceforth called polar adhesion, promotes the order of collective motion \cite{Coates:2001, Siu:2011}. In contrast, irregular polarization in the concentration of adhesion molecules may lead to disordered motion. In fact, a recent study reports irregular strengthening of polar adhesion as a candidate origin of cell aggregation \cite{Matsumoto:2024}, which poses difficulties in treating neuronal injury \cite{Johnson:2005}. In this study, we aim to provide a simple example of the effects of irregular polarization on adhesion.

As a model case of irregular polarization, we consider adhesion at the rear edge as shown in Fig.~\ref{fig:concept}(b). We call this polarized adhesion ``antipolar'' adhesion, which means that its adhesive site is opposite to the site of polar adhesion in Fig.~\ref{fig:concept}(a), with the direction of motility $\bm p$. 
Based on the comparison between polar and antipolar adhesion, we explore the effects of irregular polarization of adhesion sites on the directional order and disorder in the cell motility.

Further, we consider the choice of an appropriate adhesion type to construct a physical model of antipolar adhesion. 
Cell-cell adhesion is typically classified as homophilic adhesion or heterophilic adhesion \cite{Alberts:2022}. Homophilic polarized adhesion is well known for the DdCad1 adhesion molecule of {\it Dictyostelium discoideum} (dicty) and strongly induces cellular alignment, which stabilizes the order of motion \cite{Beug:1973, Muller:1978, Matsushita:2017}. However, the cellular alignment makes it difficult for us to observe the pure effect of antipolar adhesion. Hence, we investigate another candidate adhesion type. 

We consider a model cellular system using heterophilic adhesions. Heterophilic adhesion only stabilizes the cellular contact between different adhesion molecules. Many types of adhesion molecules are well known, including L-selectin for leukocytes \cite{Luster:2005} and tgr for dicty \cite{Wang:2000}, which guide cell migration. Using this feature, we employ an adhesion molecule that is not polarized, namely, isotropic adhesion sites, as shown in Fig.~\ref{fig:concept}(c). The isotropy suppresses the aforementioned specific cellular alignment \cite{Matsushita:2024}.
To model this case, we consider that one of the two molecules participating in the heterophilic adhesion, herein called molecule A, forms a polar adhesion site, as shown in Fig.~\ref{fig:concept}(a) or an antipolar adhesion site, as shown in Fig.~\ref{fig:concept}(b). The second adhesion molecule, referred to as adhesion molecule B, is an isotropic adhesion site, as shown in Fig.~\ref{fig:concept}(c).

The current study examines the differences between polar and antipolar adhesion in terms of the spontaneous directional order of motility among cells. 
We construct a model for these adhesions using the cellular Potts model \cite{Graner:1992,Glazier:1993,Savill:1997,Anderson:2007,Scianna:2013}, which is easy to implement for both polar and antipolar adhesions \cite{Zajac:2003,Matsushita:2017}. We find that antipolar adhesion induces collective motility ``disorder'', which is a disorder in the directions of the cells, hence  inhibiting collective migration. As a result, antipolar adhesion leads to a sub-diffusive state in the whole cellular system, which may be similar to the migration-inhibited state, such as in epithelial cell systems \cite{Bi:2016,Saito:2024,Marzio:2025}. 
The effect of antipolar adhesion appears at a low motility strength $\varepsilon$ and subsequently results in a depinning transition to a collective migration state with increasing $\varepsilon$, as schematically shown in Fig.~\ref{fig:concept}(d). Motility disorders can be explained by a cooperative effect between antipolar adhesion and motility persistence in the mean-field approach. In conclusion, we propose that antipolar adhesion provides a candidate example of irregular interactions that cause a collective motility disorder.

\section{\label{sec:model}Model and Simulation}

We employ the cellular Potts model on a square lattice \cite{Graner:1992}. To easily observe the directional order of motility, we also impose periodic boundary conditions, in which the order results in high values of its order parameter. In this model, the Potts states represent the cell configuration; $m(\bm r)$ represents the Potts state at site $\bm r$, and $m(\bm r)$ can take an integer value from 1 to $N$. It indicates that the cell with an index of $m(\bm r)$ occupies the site. This representation indicates that the domain of state $m(\bm r)$ = $m$ represents the shape of the $m$th cell. In this study, we consider only the confluent state, which inhibits complex cluster formation arising from cell aggregation, to simplify our observations. Here, the confluent state indicates that the cells cover all the sites on the square lattice.

The model simulates the time series of cell configuration using the Monte Carlo method. 
The simulation generates a configuration for each Monte Carlo step. The Monte Carlo step (MCS) consists of $16L^2$ copies of the state. The state copy is the copy of a randomly chosen site $\bm r$ from its randomly chosen neighbor site $\bm r'$. The neighboring sites consist of the nearest and second-nearest sites. The state copy has the Metropolis probability of acceptance $\min$[1, $P'/P$], where $P$ and $P'$ represent the occurrence probability of the states before and after the copy, respectively.

The occurrence probability is proportional to the Boltzmann weight, $\exp[-\beta H(\{m(\bm r)\},\{\bm p_m\})]$, where $\{m(\bm r)\}$ represents the set of Potts states in the lattice, and $\{\bm p_m\}$ represents the set of leading-edge directions of the cells, as shown in Figs.~\ref{fig:concept}(a)--\ref{fig:concept}(c). In $\{\bm p_m\}$, the unit vector $\bm p_m$ represents the direction of the leading edge of the $m$th cell. The Hamiltonian ${\cal H}$ consists of three terms, as below
\begin{align}
    {\cal H} = {\cal H}_{\rm adh} + {\cal H}_{\rm dri} + {\cal H}_{\rm vol}.
    \label{eq:Hamiltonian}
\end{align}
All these terms have parameters independent of $m$ and, therefore, represent a homogeneous system.

\begin{figure}[t]
\begin{center}
\includegraphics[width=1\linewidth]{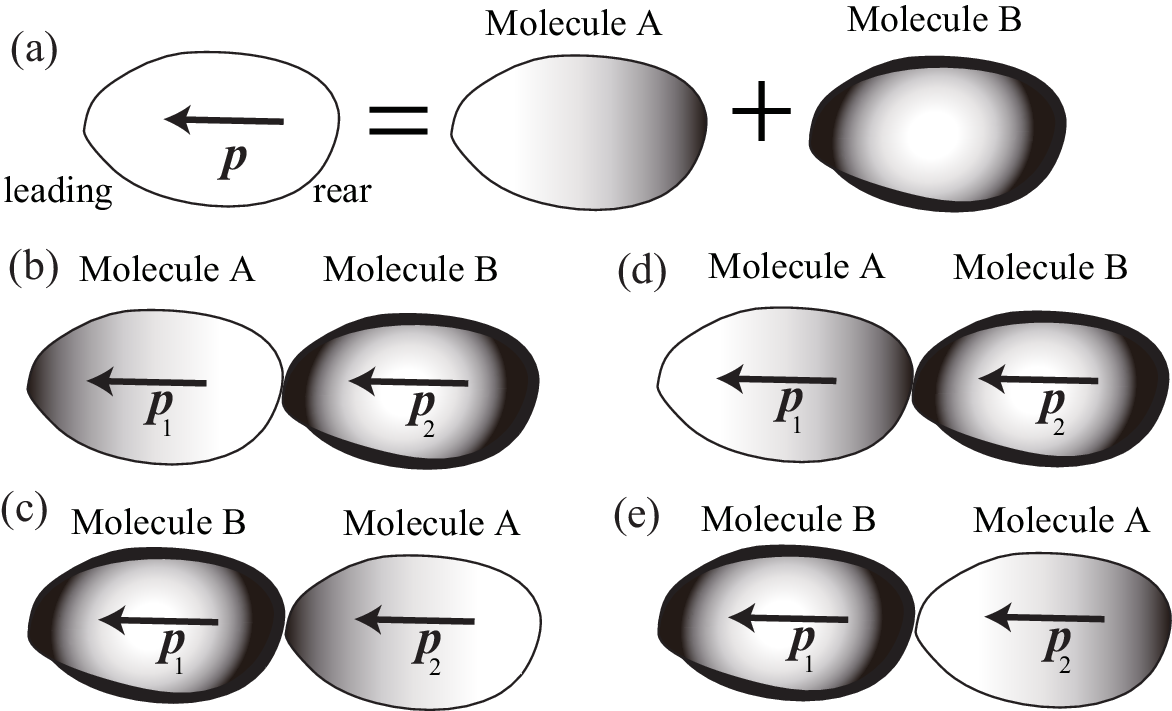}
\caption{  (a) Schematic view of the model cell corresponding to the adhesion term in Eq.~\eqref{eq:adhesion}. Dark gray-colored regions indicate high concentrations of heterophilic adhesion molecules. The concentrations of adhesion molecules A and B are assumed in (a). The arrow $\bm p$ represents the motility direction of the cell. Two cells (m = 1 and m = 2) are shown aligning in the adhesion polarization direction for polar ((b) and (c)) and antipolar adhesion ((d) and (e)). (b) shows both molecule A on the surface of cell 1 and molecule B on the surface of cell 2 for polar adhesion. (c) shows molecule B on the surface of cell 1 and molecule A on the surface of cell 2 for polar adhesion. (d) is the case of antipolar adhesion corresponding to (b), and (e) is the case of antipolar adhesion corresponding to (c). In the panels, arrows $\bm p_1$ and $\bm p_2$, respectively, represent the motility directions of cell 1 and cell 2.}
\label{fig:model_for_adhesions}
\end{center}
\end{figure}

The first term of the right-hand side in \eqref{eq:Hamiltonian} represents cell-cell adhesion with polarization of the molecular concentration  \cite{Zajac:2003, Vroomans:2015, Matsushita:2017, Matsushita:2018}. The term is concretely
\begin{align}
    {\cal H}_{\rm adh} = \sum_{\bm r \bm r'}\left\{\gamma_0 - \gamma \left[\xi^A_{m(\bm r)}(\bm r) + \xi^A_{m(\bm r')}(\bm r)\right]\right\} \nonumber \\
    \times \left(1-\delta_{m(\bm r)m(\bm r')}\right). \label{eq:adhesion}
\end{align}
Here, the summation over site pairs $\bm r\bm r'$ is performed on the neighboring sites defined above. $\gamma_0$ and $\gamma$ are, respectively, the isotropic and polarized parts of the surface tension between cells. The surface tensions represent the cells with the adhesion molecule A at a polarized concentration and the isotropic adhesion molecule B, as shown in Fig.~\ref{fig:model_for_adhesions}(a).
In our simulation, $\gamma_0$ is a constant. $\gamma$ can be either positive or negative, corresponding to polar and antipolar adhesion, respectively.  $\xi_m(\bm r)$ is the tension reduction contribution due to molecule A for site $\bm r$ on the periphery of the $m$th cell \cite{Matsushita:2017, Matsushita:2024}
\begin{align}
\xi^A_m(\bm r) = 1- \bm e_m(\bm r) \cdot {\bm p_m}.
\end{align}
Here, $\bm e_m(\bm r)$ is the unit vector from the center of the $m$th cell $\bm R_m$ to site $\bm r$. The equation indicates that adhesion contact at the leading edges between the two cells gains surface tension energy. The contribution of molecule B is unity owing to its isotropy.  

We estimate the difference in the effects between polar and antipolar adhesions to design the simulation plan from Eq.~\eqref{eq:adhesion}. For this estimation, we assume a two-cell configuration, as shown in Figs.~\ref{fig:model_for_adhesions}(b-e). Cells 1 and 2 make head-to-tail contact in the same direction as their motility. For polar adhesion, the contact shown in Fig.~\ref{fig:model_for_adhesions}(b) can not gain energy, since the high-concentration regions of the two cell surfaces are mismatched. Therefore, the combination of molecule A in cell 1 and molecule B in cell 2 does not stabilize the direction order of cell motility. In contrast, the combination of molecule B in cell 1 and molecule A in cell 2, as shown in Fig.~\ref{fig:model_for_adhesions}(c), stabilizes the order of motility. The order drives the overtaking of cell 2 to gain energy through adhesion. In fact, as a result of this overtaking in the periodic system, empirical simulations show that cells exhibit an ordered motion \cite{Matsushita:2018,Matsushita:2021a}. 

Unlike polar adhesion, the effects of antipolar adhesion are not well known.
To speculate on the effect, we consider a similar two-cell configuration, as shown in Fig.~\ref{fig:model_for_adhesions}(d) and \ref{fig:model_for_adhesions}(e). The cellular contact shown in Fig.~\ref{fig:model_for_adhesions}(d) can gain energy, since the high-concentration regions of the two cell surfaces match and induce a directional order in motility. In contrast, the contact shown in Fig.~\ref{fig:model_for_adhesions} (e) is mismatched. As a simple speculation, based on the similarity of polar adhesion, the combination of molecule A in cell 1 and molecule B in cell 2 drives the backward overtaking of cell 1. Here, backward overtaking means that cell 1 moves around cell 2 in the direction opposite to its motility. Based on empirical simulations of polar adhesion, antipolar adhesion may generate a driving force in the direction opposite to that of $\bm p_m$. Therefore, antipolar adhesion may drive backward motion order as a motility effect, in contrast to polar adhesion.

From these discussions, we select the design of our simulation to sweep $\gamma$ from negative to positive values. Collective motion is associated with the direction of motility when $\gamma$ $>$ $0$. We expect that the collective motion would occur in the direction opposite to that of motility for the case of $\gamma$ $<$ $0$. Additionally, with increased motility strength $\varepsilon$ $>$ 0, the direction-change value of collective motion in $\gamma$ shifts to a negative value of $\gamma$ in the phase space, since the motility stabilizes the collective motion in the direction of motility \cite{Matsushita:2019}.  The $\gamma$-sweep simulation with self-motility control of the cells effectively examines this expectation for directional changes in collective motion.

To perform the simulation, we employ self-motility of cells, represented by the second term of r.~h.~s. in Eq.~\eqref{eq:Hamiltonian} \cite{Rappel:1999, Kabla:2012, Matsushita:2020a, Matsushita:2022a}
\begin{align}
{\cal H}_{\rm drv} = -\varepsilon \sum_{\bm r}\xi_{m(\bm r)}(\bm r), \label{eq:hamiltonian_motility}
\end{align}
where $\xi_m(\bm r)$ coincides with $\xi^A_m(\bm r)$, since the direction of adhesion edge is the motile direction $\bm p_m$ or the opposite direction of $\bm p_m$. Hence, we set $\xi_m(\bm r)$ = $\xi_m^A(\bm r)$. The opposite case can represent a negative value of $\gamma$. $\varepsilon$ is the strength of this motility and employs a positive value owing to its consistency with the direction of motility. This term is similar to the $\gamma$-proportional part of ${\cal H}_{\rm adh}$ in Eq.~\eqref{eq:adhesion} and differs from ${\cal H}_{\rm adh}$ in its summation over the sites.  In the confluent system discussed here, we assume that the polarized adhesion in Eq.~\eqref{eq:adhesion} for $\gamma$ $>$ 0 has a similar effect as that of motility in Eq.~\eqref{eq:hamiltonian_motility} when we ignore the details of cell shapes on average in homogeneous states. Therefore, although we clearly distinguish the two terms in the simulation for the microscopic levels in \S~\ref{sec:result}, we identify the terms later for the macroscopic state of the order parameter as a result of the coarse-graining in the mean field approximation in \S~\ref{sec:mean_field}.

\begin{figure*}[t]
\begin{center}
\includegraphics[width=1\linewidth]{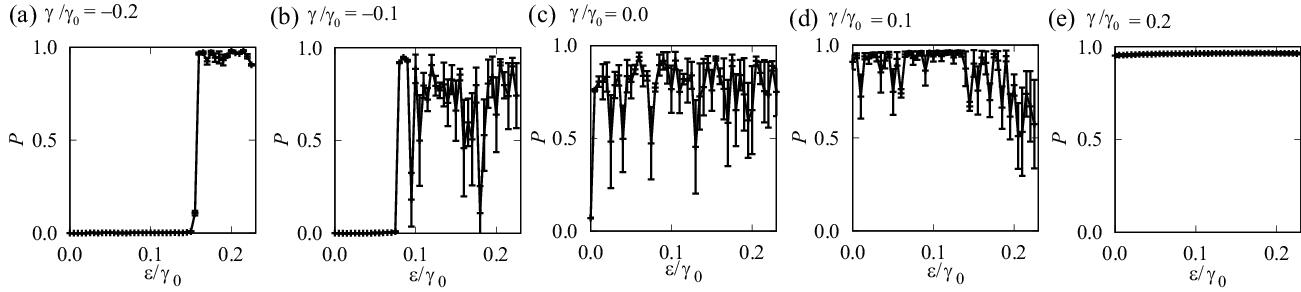}
\caption{ The order parameter $P$ as a function of the strength of motility  $\varepsilon/\gamma_0$ for (a) $\gamma/\gamma_0$ = -0.2, (b) $\gamma/\gamma_0$ = -0.1, (c) $\gamma/\gamma_0$ = 0.0,  
(d) $\gamma/\gamma_0$ = 0.1, and  (e) $\gamma/\gamma_0$ = 0.2. The error bars show the deviation from the average over time and indicate fluctuations in the time course of $P$. }
\label{fig:orderparameter_for_adhesions}
\end{center}
\end{figure*}

The third term of r.~h.~s. in Eq.~\eqref{eq:Hamiltonian} represents the empirical restriction of cell area $A_m$ = $\sum_{\bm r }\delta_{mm(\bm r)}$ \cite{Graner:1992}
 \begin{align}
    {\cal H}_{\rm vol} = \kappa A_0 \sum_m\left(1-\frac{A_m}{A_0}\right)^2, \label{eq:volume}
\end{align}
where $A_0$ is the restriction target area of the cell.

The unit motility vector of the $m$th cell $\bm p_m$ induces  a random cell walk  with a certain  persistent length \cite{Li:2008,Takagi:2008, Sadhukan:2025} by following the equation of motion\cite{Szabo:2006, Kabla:2012} 
\begin{align}
\frac{d}{dt} \bm p_m(t) =  \frac{1}{\tau}\left(\hat I - \hat P(\bm p_m)\right) \frac{\bm d_m}{a}. \label{eq:motility}
\end{align}
Here, $\tau$ is the ratio of the motility relaxation time to the relaxation time of $\bm d_m(t)$ and determines the persistence length of the cell motion.
$\bm d_m(t)$ is the displacement vector of the cell center $\bm R_m$ at time $t$. $a$ is the lattice constant, $\hat I$ is the unit matrix, and $\hat P(\bm p_m)$ is the projection matrix in the direction of $\bm p_m$. Equation~\eqref {eq:motility} expresses that $\bm p_m$ relaxes in the direction of displacement $\bm d_m$. The equation is integrated for each MCS with the Euler method with time difference $\Delta t$, and then $\bm d_m$ is updated as a displacement of the cell center position $\bm R_m$ = $\sum_{\bm r} \bm r\delta_{m(\bm r)m}/\sum_{\bm r}\delta_{m(\bm r)m}$.

Since we consider a confluent state with high cell density,  the high density transiently causes jamming of cells that depends on the initial state. Since transient jamming might lead to the misidentification of a collective motility disorder owing to adhesion at a steady state, a short relaxation time is preferable. To achieve a short relaxation time, we empirically choose $\beta$ = 0.2, $\gamma_0$ = 4.0, $\kappa A_0$ = 1.0 with $A_0$ = 64. In our simulation, by setting $\chi$, we employ the directional order of motility for $\varepsilon$ $>$ 0 without additional polarization $\gamma$, in order to examine whether antipolar adhesion inhibits the order. For this, we use  $\tau/\Delta t$ = 5.0, which empirically realizes the ordered motility at $\gamma$ $=$ $0$, owing to the persistence in cell motility \cite{Kabla:2012, Matsushita:2019, Matsushita:2020a, Bertrand:2024} similar to self-propelled systems\cite{Peruani:2006, Baskaran:2008, Deseigne:2010, Weber:2013, Lober:2015, Hiraoka:2016, Ohta:2017}. To sufficiently reduce the finite-size effects under periodic boundary conditions in the confluent state, we empirically employ $L$ = 192 and $N$ = $L^2/V_0$ = 576 as a tractable system size. 

Our simulation starts from an initial square lattice configuration of cells with randomly oriented motility directions.
To reduce the dependence of observations on the initial state, stationary cell states should be obtained. Therefore, we perform a sufficiently long relaxation of the initial state using Monte Carlo simulations. The number of relaxation MCS in the migrating cells of $\varepsilon$ $>$ 0 or $\gamma$ $>$ 0 is empirically in the order of $10^3$--$10^4$ MCS, and we carry out relaxation during $t_s$ = $10^5$ MCS; then observe the order in motility and collective velocity of the motion.
Based on this observation, we calculate the order parameter of motility $\bm p_m$ \cite{Vicsek:1995}:
\begin{align}
    P = \left|\frac{1}{(t_e-t_s)N}\sum_m^N \int_{t_s}^{t_e} \bm p_m(t) dt\right|.
\end{align}
Using the order parameter, we confirm the presence of a directional order in motility. We obtain the time average for $t_e$ = 5$\times$10$^5$ MCS.
Further, we confirm the existence of collective motion by calculating the norm of collective velocity,
\begin{align}
    v = \left|\frac{1}{(t_e-t_s)N}\sum_m^N \int_{t_s}^{t_e} \bm d_m(t) dt\right|.
\end{align}
When $v$ = 0 under the periodic boundary condition at motility strength $\varepsilon$ $>$ 0, it implies a collective cellular motility disorder. Therefore, we can probe the disordered state through the observation of $v$ at finite strengths of motility $\varepsilon$.

\begin{figure*}[t]
\begin{center}
\includegraphics[width=1\linewidth]{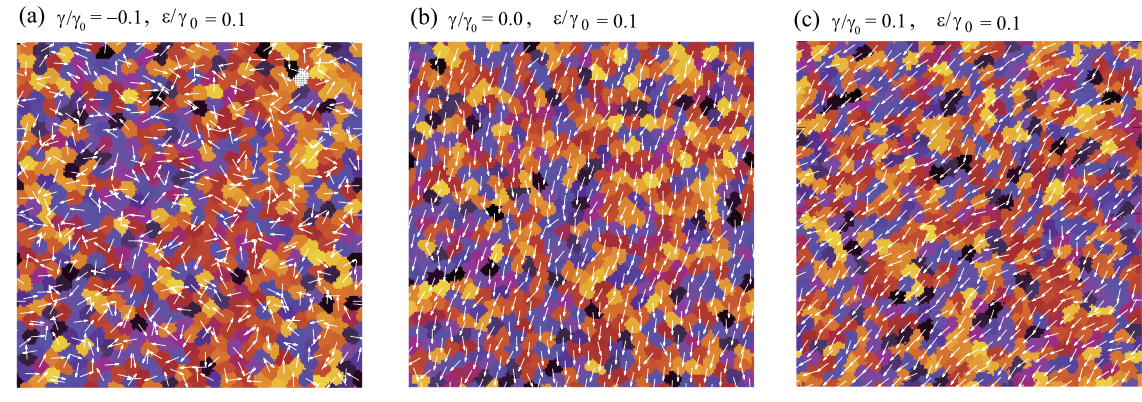}
\caption{ Snapshots of states at $t$ = $t_e$ for (a) $\gamma/\gamma_0$ = -0.2, (b)  $\gamma/\gamma_0$ = 0.0, and (c) $\gamma/\gamma_0$ = 0.2. For motility, we take $\varepsilon/\gamma_0$ = 0.1 (see text). The colored domains represent individual cells, and differences in color indicate distinct cells. The white arrows represent the direction of motility.}
\label{fig:snapshots}
\end{center}
\end{figure*}
\begin{figure*}[t]
\begin{center}
\includegraphics[width=1\linewidth]{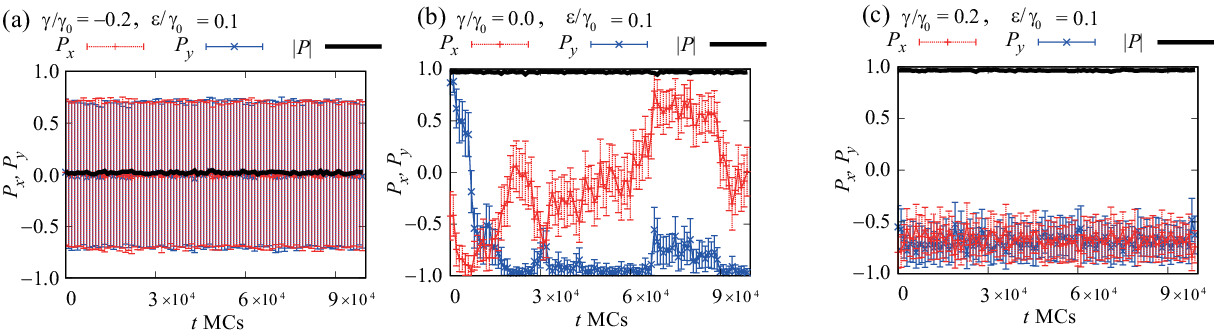}
\caption{ Time course of the components of average motility over cells for (a) $\gamma/\gamma_0$ = -0.2, (b)  $\gamma/\gamma_0$ = 0.0, and (c) $\gamma/\gamma_0$ = 0.2. For motility, we take $\varepsilon/\gamma_0$ = 0.1, as in Fig.~\ref{fig:snapshots}(a-c). The $+$ symbol with dashed error bars and $\times$ symbol with solid error bars represent $P_x$ and $P_y$, respectively. The thick line represents $|P|$ = $\sqrt{P_x^2+P_y^2}$.}
\label{fig:components-order}
\end{center}
\end{figure*}
\begin{figure}[t]
\begin{center}
\includegraphics[width=0.9\linewidth]{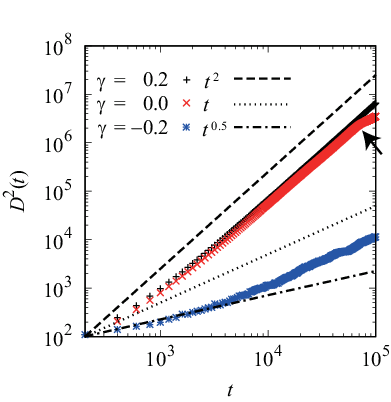}
\caption{
Mean square displacement $D^2(t)$ for  $\gamma/\gamma_0$ = $-$0.2 ($+$), 0.0 ($\times$), 0.2 (+\llap{$\times$}) with $\varepsilon/\gamma_0$ = 0.1. Dashed, dotted, and dashed-dotted lines represent $t^2$, $t$, and $t^{1/2}$ lines. 
The arrow indicates the crossover point from the ballistic to the diffusive behavior for $\gamma$ = 0.0.
}
\label{fig:mean-squere-displacement}
\end{center}
\end{figure}

\section{Results}\label{sec:result}
We analyze the order parameters as functions of $\gamma$ to explore the typical differences between polar and antipolar adhesion.  
For this analysis, we calculate the dependence of the order parameter $P$ on the motility strength $\varepsilon$ for various values of $\gamma$. The calculation results are plotted in Figs.~\ref{fig:orderparameter_for_adhesions}(a-e); the antipolar cases are in panels (a) and (b), the isotropic case is in panel (c), and the polar cases are in panels (d) and (e).

In antipolar adhesions for $\gamma$ $<$ 0, as shown in Figs.~\ref{fig:orderparameter_for_adhesions}(a) and \ref{fig:orderparameter_for_adhesions}(b), the order parameter $P$ at $\varepsilon$ = 0 indicates a commonly disordered state. As $\varepsilon$ increases, they exhibit an abrupt transition from the disordered state to fluctuating ordered states. We define the fluctuating ordered state as a finite value of $P$ with fluctuations, indicated by error bars calculated as deviations from the time average. We then denote the transition points  $\varepsilon_d$, whose values depend on the polarization strength in adhesion $\gamma$, and decrease as $\gamma$ increases. For isotropic adhesion at $\gamma$ = 0, $\varepsilon_d$ is 0, and only the fluctuating ordered state remains for $\varepsilon$ $>$ 0, as shown in Fig.~\ref{fig:orderparameter_for_adhesions}(c).

For the cases of polar adhesion shown in Fig.~\ref{fig:orderparameter_for_adhesions}(d) and \ref{fig:orderparameter_for_adhesions}(e), the order-parameter fluctuation is observed at least up to $\gamma$ = 0.2$\gamma_0$. The fluctuation then becomes small, and $P$ is almost 1 for $\gamma$ values higher than $\gamma$ = 0.2$\gamma_0$. 
Here, we define a stable ordered state as a state with $P$ approximately 1.
This observation implies that a surface tension energy barrier exists and changes the state from fluctuating to stable order in the range from $\gamma$ = 0.1$\gamma_0$ to $\gamma$ = 0.2$\gamma_0$.
For a more detailed clarification of this state change, we would have to calculate the detailed $\gamma$-dependence of the order parameter. Therefore, we leave the clarification of the change for later, in terms of the calculation of $\gamma$-dependence. Before the calculation, we confirm the motility order of these states from the configurations of cells $\{m(\bm r) \}$ and their motility vectors $\{\bm p_m\}$.

From the order parameter, at least, we expect the existence of the following three states: the disordered state $P$ = 0, $P$ $>$ 0 with its time fluctuations, and $P$ $\simeq$ 1 with almost no time fluctuation. 
The states are reflected in the directions of motility from the snapshot configurations.
To further understand the interpretations of these states, the snapshot configurations are examined. In particular, the observation may be effective for interpreting fluctuations in ordered states.
Figures~\ref{fig:snapshots}(a-c) show the snapshots of cell configuration and motilities for $\gamma=-0.2\gamma_0$, $\gamma=0$, and $\gamma=0.2\gamma_0$. We employ $\varepsilon$ = 0.1$\gamma_0$ to get the configuration of the disordered, fluctuating-ordered, and stable-ordered states shown in panels \ref{fig:snapshots}(a), \ref{fig:snapshots}(b), and \ref{fig:snapshots}(c), respectively. In the disordered state in Fig.~\ref{fig:snapshots}(a), the configuration of motility indicated by the white arrows is random. In contrast, the two ordered states show ordered motility in Fig.~\ref{fig:snapshots}(b) and Fig.~\ref{fig:snapshots}(c). The motility configurations are consistent with the value of $P$ in the disordered and ordered states. In particular, the fluctuating-ordered state is clearly ordered. Thus, $P$ in the fluctuating-ordered state indicates the time fluctuations in the direction of ordered motility in its time course.

To further examine the directional fluctuation of the order parameter corresponding to the snapshots in Fig.~\ref{fig:snapshots}(a-c), we calculate the average value of the order parameter component $P_\alpha$ for $\alpha$ = $x$ or $\alpha$ = $y$ for the three states at $\varepsilon$ = 0.1$\gamma_0$. 
\begin{align} 
P_\alpha(t) = \frac{1}{N}\sum p_{m,\alpha}(t).
\end{align}
Here, $p_{m,\alpha}$ is the $\alpha$-component of $\bm p_m$. 
We show the time course of these components in Fig.~\ref{fig:components-order}(a-c) with error bars, which indicate deviation among the cells. 
For the disordered state at $\gamma$ = $-$0.2$\gamma_0$, the components, $P_x$ and $P_y$, are zero, and the deviations among cells are comparably larger than those for ordered states, as shown in Fig.~\ref{fig:components-order}(a). This reflects the disordered configuration of cell motility directions.

The components for the ordered motility states are shown in Figs.~\ref{fig:components-order}(b) and \ref{fig:components-order}(c), respectively, for the fluctuating ordered state at $\gamma$ = 0.0 and stable ordered state at $\gamma$ = 0.2.  The motility disorder among cells in these states is smaller than that in the disordered state, as shown in Fig.~\ref{fig:components-order}(a). The fluctuating ordered state exhibits drift motion in components $P_x$ and $P_y$ as shown in Fig.~\ref{fig:components-order}(b).  This contrasts with the stable ordered state, which maintains constant component values, as shown in Fig.~\ref{fig:components-order}(c). Furthermore, the absolute value of $|P|$ remains nearly unity over time. The result clearly indicates that the fluctuation in the order parameter $P$ shown in Figs.~\ref{fig:orderparameter_for_adhesions}(a-d) originates not from the destabilization of motility order among cells but from the directional fluctuation in motility order. 

Based on this understanding of the three states, we shift our focus to the features of their collective motion. Collective motion is expected to be either diffusive or ballistic when averaged over the cells. To examine this, we consider the mean square displacement of the collective coordinates of the cells, as follows:
\begin{align}
D^2(t) = \frac{1}{N}\left|\int_{t_s}^{t} \sum_m {\bm d}_{m}(t) dt\right|^2.
\end{align}
For the disordered state at $\gamma$ = $-0.2\gamma_0$, $D^2(t)$ is on the line $t^{1/2}$ in the short time up to 3 $\times$ 10$^3$ MCS. In this short time, the collective motion is sub-diffusive \cite{Angelini:2011, Wurthner:2026}, which reflects a disordered state, in which the motility directions of cells are random. 
With increasing $t$, $D^2(t)$ exhibits a crossover from $t^{1/2}$ to $t$, indicating a diffusive motion. 

For the fluctuating ordered state at $\gamma$ = 0.0 and the stable ordered state at $\gamma$ = 0.2$\gamma_0$, $D^2(t)$ commonly exhibits the function form of $t^2$, indicating a ballistic motion, in the short time up to 10$^4$ MCS. With increasing $t$, $D^2(t)$ in the fluctuating ordered state at $\gamma$ = 0.0 exhibits a crossover from ballistic motion to diffusive motion. Within the time range of our observation, the ballistic motion persists in the stable ordered state at $\gamma$ = 0.2$\gamma_0$. In contrast, the fluctuating ordered state exhibits a crossover from ballistic to diffusive motion in the time range of $10^4$ MCS.

Based on the behavior of $D^2(t)$, the states correspond to cell motions as follows: the disordered state consists of non-migrating cells for a certain short time, the fluctuating ordered state is a persistent random walk, and the stable ordered state is a ballistic motion. In particular, this specification of collective migration indicates that the abrupt state change observed in Figs.~\ref{fig:orderparameter_for_adhesions}(a) and
\ref{fig:orderparameter_for_adhesions}(b) is a depinning transition from a non-migrating state to a collective migration state. In addition, the non-migrating state corresponds to a directional disorder of motility, which is referred to as a collective motility disorder in the present study. 

\begin{figure}[t]
\begin{center}
\includegraphics[width=0.9\linewidth]{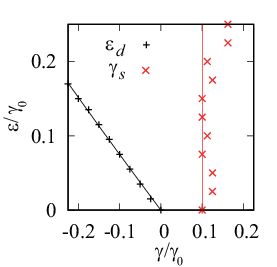}
\caption{
Phase diagram from simulation. $\varepsilon_d$ represents the depinning transition point in $\varepsilon$ as a function of $\gamma$, which is defined as the abruptly changing point of $P$. $\gamma_s$ represents the stabilization point, which is defined by the $\gamma$-value at which $D^2(t)$ does not exhibit diffusion motion in our simulation time range for values of $\gamma$ larger than $\gamma_s$ for each $\varepsilon$.
}
\label{fig:phase-diagram}
\end{center}
\end{figure}

Our observations thus far indicate that three states arise depending on whether the adhesion is polar or antipolar. Next, we clarify the phase diagram of these systems and determine the depinning transition point from disordered motility to the ordered state or state-crossover regions between the ordered states. As a starting point for this clarification, we specify the disordered state via the order parameter and determine the transition points $\varepsilon_d$ from the disordered state to the fluctuating ordered state. To do this, we calculate the transition points as a function of $\gamma$ and plot them with $+$ symbols in Fig.~\ref{fig:phase-diagram}.
\begin{figure}[t]
\begin{center}
\includegraphics[width=0.9\linewidth]{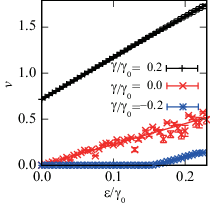}
\caption{
Collective velocity $v$ as a function of $\varepsilon$.
The $+$ symbol, $\times$, and $+$\llap{$\times$} are data for $\gamma/\gamma_0$ = 0.2, 0.0, and -0.2, respectively. 
}
\label{fig:collective velocity}
\end{center}
\end{figure}
The transition point $\varepsilon_d$ is negatively proportional to $\gamma$, as shown in the equation below;
\begin{align}
    \varepsilon_d(\gamma) = -k \gamma. \label{eq:jam_points}
\end{align}
The value of $k$ is approximately 3/4. 

Next, we examine the state-change conditions from a fluctuating to a stable ordered state. To achieve this, we use the existence of a crossover from ballistic to diffusion motion in $D^2(t)$. We calculate $D^2(t)$ with varying $\gamma$ for the values of $\varepsilon/\gamma_0$ = 0.025, 0.050, \dots, 0.250.
We examine the lowest value of $\gamma$ where the diffusion motion is absent and define the lowest value as the stabilization point $\gamma_s$ as a function of $\varepsilon$. $\gamma_s(\varepsilon)$ is plotted by $\times$ symbols in Fig.~\ref{fig:phase-diagram}. The value of $\gamma_s$ is approximately 0.1 at least for $\varepsilon$ larger than 0.2. Therefore, stabilization is understood to be purely an effect of the tension-energy barrier due to adhesion, independent of motility. 

Next, we examine the collective velocity of the polar and antipolar adhesions. 
For polar adhesion at $\gamma$ = 0.2$\gamma_0$, the collective velocity $v$ has a finite value even at $\varepsilon$ = 0; this is the guiding effect of the polar adhesions \cite{Matsushita:2018}. We find that $v$ increases almost linearly with $\varepsilon$.
For the isotropic case at $\gamma$ = 0, $v$ becomes 0 at $\varepsilon$ = 0. $v$ is almost proportional to $\varepsilon$, and small deviations from linear dependence do appear, in contrast to the case of polar adhesion. For antipolar adhesion at $\gamma$ = $-$0.2, the absence of migration due to the disordered state results in the finite threshold values for finite values of $v$. The threshold value is equal to the phase boundary $\varepsilon_d$ in Fig.~\ref{fig:phase-diagram}. The value of $v$ increases linearly with $\varepsilon$ above $\varepsilon$ = $\varepsilon_d$. Therefore, the depinning transition occurs as an onset of collective motion at $\varepsilon_d(\gamma)$. The dependencies of $v$ on $\varepsilon$ are consistent with the state observed via the order parameter and $D^2(t)$.

\section{Mean Field Approach} \label{sec:mean_field} 

In our simulation, antipolar adhesion is found to result in a collective motility disorder for small $\varepsilon$.
The emergence of the disordered state could be explained using the mean-field approach. In this section, we consider solutions to obtain an intuitive understanding of this phenomenon.
In particular, we focus on the reason underlying the collective motion not occurring in the direction opposite to motility, as expected in the discussion in \S~\ref{sec:model}. Using this approach, the mechanism is determined based on the persistent random walk of the cells. 

We consider a simple mean field model based on the assumption of a driving force
\begin{align}
\bm d_m = - \frac{d H_{\rm dri}}{d {\bm R}_m}  - \frac{d H_{\rm adh}}{d \bm R_m}. \label{eq:driving_assumption}
\end{align}
Here, we neglect the volume term in Eq.~\eqref{eq:volume}, since the term does not affect the order of motility, the stability of which is discussed here. 
Moreover, we neglect the isotropic adhesion part, which cannot be the origin of the order of motility, since it cannot enable cells to inform other cells of the motility direction.
Further, we assume a persistent random walk of cells, as described by Eq.~\eqref{eq:motility}.

 We consider the velocity of the collective coordinates, $\sum_m {\bm d}_m/N$. To achieve this, we average Eq.~\eqref{eq:driving_assumption} for all the cells.
 As discussed previously \cite{Matsushita:2018}, we suppose that the first term of Eq~\eqref{eq:driving_assumption} is approximately $-\gamma \bm p_m \cdot \bm R_m$ + const for each cell in confluence. This is because the confluent state enables cells to adhere to the surrounding cells and exert a driving force due to polarization in adhesion-molecule concentration. Therefore, the term produces the following driving force from the polarization in adhesion averaged over cells
\begin{align}
\frac{1}{N}\sum_m\left<\frac{d H_{\rm adh}}{d \bm R_m}\right> = - c_{\gamma}\gamma \bm P, 
\end{align}
Here, $\bm P$ = $\left<\sum_m \bm p_m/N\right>$ and the bracket is the time average. $c_{\gamma}$ is a response coefficient of the linear response for cell configuration to the polarization in adhesion. 

We assume that the proportional part of motility in the second term of Eq.~\eqref{eq:driving_assumption} is approximately $-\epsilon  \bm p_m \cdot \bm R_m$ for each cell, and therefore, the second term is \cite{Szabo:2006,Kabla:2012}
\begin{align}
    \frac{1}{N}\sum_m\left<\frac{d H_{\rm dri}}{d {\bm R}_m}\right> = - c_{\varepsilon}\varepsilon \bm P
\end{align} in time-averaged ordered motility; $c_{\varepsilon}$ is a response coefficient of the linear response for cell configuration.
Therefore, by averaging Eq.~\eqref{eq:driving_assumption} over time as follows, we have
\begin{align}
\bm v \simeq (c_{\varepsilon}\varepsilon + c_{\gamma}\gamma) \bm P, \label{eq:mean_field_eq}
\end{align}
in the linear response regime, in which the contributions of adhesion and motility are assumed to be independent.
Here, $\bm v$ = $\left<\sum_m\bm d_m/N\right>$.

For the case of $-c_{\gamma}\gamma$ $<$ $c_{\varepsilon}\varepsilon$, solutions of finite norms, $|\bm v|$ $>$ 0 and $|\bm P|$ $>$ 0, are an acceptable solution for Eq.~\eqref{eq:mean_field_eq}.
In contrast, for $-c_{\gamma}\gamma$ $\geq$ $c_{\varepsilon}\varepsilon$, the solution should be 
\begin{align}
    \bm v \cdot \bm P \leq 0. \label{eq:steady_state_for_driving}
\end{align}
$\bm v$ should be in the anti-parallel direction of $\bm P$. The result is consistent with our conclusion from the intuitive discussion in \S~\ref{sec:model}.

Assuming a persistent random walk from Eq.~\eqref{eq:motility}, $\bm p_m$ is aligned in the direction of cell displacement $\bm d_m$. The effects of the alignment on motility lead to a stable state, where $\bm v$ is in a direction parallel to $\bm P$ \cite{Weber:2013,Hanke:2013,Hiraoka:2016,Hiraoka:2017,Matsushita:2019}. 
Namely, 
\begin{align}
 \bm v \cdot \bm P \geq 0. \label{eq:steady_state_for_motility}
\end{align}
The equation contrasts with a simple cell random walk, which imposes no special constraints on the direction of cell motility.  
If we assume the collective motion of cells, where the absolute value of $\bm v$ is finite, it would give rise to a macroscopic inconsistency between the steady states owing to the driving force in Eq.~\eqref{eq:steady_state_for_driving} and the motility dynamics in Eq.~\eqref{eq:steady_state_for_motility}. 

The discussion implies
\begin{align}
    \bm v = \bm P = \bm 0, \label{eq:jam_solution}
\end{align}
for $-c_{\gamma}\gamma$ $\geq$ $c_{\varepsilon}\varepsilon$.
There exists an instability of collective motion against a state change to collective motility disorder that may imply a non-migrating state of cells in a certain short time.
This solution in Eq.~\eqref{eq:jam_solution} does not agree with the simple expectation of backward motion order in \S~\ref{sec:model}; a collective motion does not occur in the inverse direction of ordered motility in our simulation. The mean-field approach implies that the difference from expectation originates from the motility dynamics for the persistent random walk in Eq.~\eqref{eq:motility}.

The discussion leads us to the intuitive explanation for the dependence of the depinning transition point $\varepsilon_d$ on $\gamma$. In particular, $\varepsilon_d$ values are on the line
\begin{align}
     \varepsilon_d(\gamma) = - \frac{c_{\gamma}}{c_{\varepsilon}}\gamma,
\end{align}
which agrees with the phase diagram in Fig.~\ref{fig:phase-diagram}. 
$c_\varepsilon/c_{\gamma}$ is equal to $k$ in Eq.~\eqref{eq:jam_points} and depends on the details of model construction and the parameter values.

\section{Conclusion and Remarks}

In conclusion, we investigate the effects of antipolar adhesion on collective cell motion to provide deeper insights into the role of irregular polarization in adhesion. Antipolar adhesion is found to give rise to a collective motility disorder, and thereby induces non-migrating cells for a short time under low self-motility. With increasing motility strength, the cells exhibit a depinning transition from a state consisting of non-migrating cells. This is in clear contrast to polar adhesion, which usually stabilizes and accelerates collective motion \cite{Matsushita:2018}. Since the effect is not well known in the field of collective motions, it could provide a model example of irregular intercellular interactions that pose a serious obstacle to collective phenomena observed in biophysical fields.

Discussion of the mean-field approach suggests that the migration disorder due to antipolar adhesion emerges from its cooperation with the persistence of cell motility. Persistence random walk of cells is widely observed, including dicty and epidermal cells \cite{Szabo:2006, Takagi:2008, Kabla:2012}. Adhesion polarity is also observed in dicty;  in particular, the heterophilic adhesion molecule tgr is polarized \cite{Fujimori:2019} and results in a peculiar collective migration \cite{kuwayama:2013, Hashimura:2019a, Hayakawa:2024}. Dicty is one of the most advanced model organisms and is applicable to advanced genomic engineering. When polarized adhesion can be controlled in the future, we can confirm that polarization irregularities in adhesion, including antipolar adhesion, can induce collective motility disorder. 

In our results, antipolar adhesion leads to a disorder, which is inconsistent with the spontaneous motility order among cells in Eq.~\eqref{eq:steady_state_for_driving}. As a result, a threshold motility strength $\varepsilon_{d}$ in the response of collective velocity to the motility strength appears, as shown in Fig.~\ref{fig:collective velocity}. Similarly, another type of irregular interaction that stabilizes other orders of motility is expected to induce the threshold. For example, when antiferromagnetic interactions are present during cell motility, they may induce a disordered state resulting in a similar motility response. The response to frustrating interaction effects could be observed as changes in the strength of extracellular cues.

Migration disorders caused by irregularities in the confluent state are poorly understood. It is a powerful candidate origin for abnormal neuronal aggregation \cite{Johnson:2005, Matsumoto:2024}. If this candidate is a true origin, our results would imply that aggregation itself may not be essential for the migration disorders. The aggregation may rather result from the collective motility disorders. This might provide additional insights into abnormally non-migrating cells from a physics point of view and thereby contribute to a better understanding of neuronal cell migration.

Homophilic adhesion with polarization is a potential cause for irregular adhesion \cite{Coates:2001}. As discussed in the Introduction section, the system stabilizes the positional alignment of dicty cells \cite{Beug:1973, Matsushita:2017}. Therefore, the antipolar adhesion effect may be reduced there. Whether this effect is effective in homophilic adhesion remains an issue for future studies comparing homophilic polar-polar adhesion with heterophilic polar-isotropic adhesion in detail.  

We additionally provide a comment on the stabilization of the fluctuating ordered state to the ordered state at $\gamma$ = $\gamma_s$ shown in Fig.~\ref{fig:phase-diagram}. 
A candidate for stabilization phenomena may capture the transition from a fluidically moving state \cite{Bi:2016, Saito:2024, Marzio:2025} to a cooperatively moving state.
Another candidate explanation is the effect of lattice anisotropy on the cellular Potts model \cite{Glazier:1993}. Therefore, our simulation based on the cellular Potts model could not provide a clear conclusion on whether stabilization occurs during the transition from the fluid state to the glass state. The conclusion of this stabilization, that is, the exploration of the antipolar adhesion effect, is beyond the scope of this study. The conclusion may require a simulation of the off-lattice model to avoid lattice anisotropy and is postponed to a future study.

We thank S.~Yabunaka, H.~Kuwayama, H.~Hashimura and M.~Sawada for providing the relevant information. The study was supported by JSPS KAKENHI (Grant Number 23K03342) and AMED (Grant Number JP19gm1210007).
We are supported by the facilities of the Supercomputer Center, Institute for Solid State Physics, the University of Tokyo (ISSPkyodo-SC-2025-Ca-0065 and 2026-Ca-0030).

\bibliographystyle{apsrev4-2} 

\bibliography{collective_cell_migration.bib} 

@Book{Anderson:2007,
  Title                    = {Single-Cell-Based Models in Biology and Medicine},
  Author                   = {Alexander R. A. Anderson and M. A. J. Chaplain and K. A. Rejniak},
  Publisher                = {Birkhauser Verlag AG, Basel},
  Year                     = {2007}
}

@Article{Angelini:2010,
  Title                    = {Cell Migration Driven by Cooperative Substrate Deformation Patterns},
  Author                   = {Thomas E. Angelini and Edouard Hannezo and Xavier Trepat and Jeffrey J. Fredberg and David A. Weitz},
  Journal                  = {Phys. Rev. Lett.},
  Year                     = {2010},
  Pages                    = {168104},
  Volume                   = {104}
}

@Article{Angelini:2011,
  Title                    = {Glass-like dynamics of collective cell migration},
  Author                   = {Thomas E. Angelini and Edouard Hannezo and Xavier Trepat and Manuel Marquez and Jeffrey J. Fredberg and David A. Weitz},
  Journal                  = {Proc},
  Year                     = {2011},
  Number                   = {12},
  Pages                    = {4714-4719},
  Volume                   = {108}
}

@Article{Baskaran:2008,
  Title                    = {Enhanced Diffusion and Ordering of Self-Propelled Rods},
  Author                   = {Aparna Baskaran and M. Cristina Marchetti},
  Journal                  = {Phys. Rev. Lett.},
  Year                     = {2008},
  Number                   = {26},
  Pages                    = {268101},
  Volume                   = {101}
}

@Article{Beug:1973,
  Title                    = {Dynamics of antigenic membrane sites relating to cell aggregation in Dictyostelium Discoideum},
  Author                   = {H. Beug and F. E. Katz and G. Gerisch},
  Journal                  = {J. Cell Biol.},
  Year                     = {1973},
  Pages                    = {647--658},
  Volume                   = {56}
}

@Article{Camley:2016,
  Title                    = {Emergent Collective Chemotaxis without Single-Cell Gradient Sensing},
  Author                   = {Brian A. Camley and Juliane Zimmermann and Herbert Levine and Wouter-Jan Rappel},
  Journal                  = {Phys. Rev. Lett.},
  Year                     = {2016},
  Pages                    = {098101},
  Volume                   = {116}
}

@Article{Coates:2001,
  Title                    = {Cell-cell adhesion and signal transduction during},
  Author                   = {Juliet C. Coates and Adrian J. Harwood},
  Journal                  = {J. Cell Sci.},
  Year                     = {2001},
  Pages                    = {4349--4358},
  Volume                   = {114}
}

@Article{Deseigne:2010,
  Title                    = {Collective Motion of Vibrated Polar Disks},
  Author                   = {Julien Deseigne and Olivier Dauchot and Hugues Chat\'{e}},
  Journal                  = {Phys. Rev. Lett.},
  Year                     = {2010},
  Pages                    = {098001},
  Volume                   = {105}
}

@Article{Friedl:2009,
  Title                    = {Collective cell migration in morphogenesis, regeneration and cancer.},
  Author                   = {Peter Friedl and Darren Gilmour},
  Journal                  = {Nat. Rev. Mol. Cell Biol.},
  Year                     = {2009},
  Pages                    = {445--457},
  Volume                   = {10}
}

@Article{Fujimori:2019,
  author    = {Taihei Fujimori and Akihiko Nakajima and Nao Shimada and Satoshi Sawai},
  journal   = {Proc. Natl. Acad. Sci. USA},
  title     = {Tissue self-organization based on collective cell migration by contact activation of locomotion and chemotaxis},
  year      = {2019},
  number    = {10},
  pages     = {4291--4296},
  volume    = {116},
}

@Article{Glazier:1993,
  Title                    = {Simulation of the differential adhesion driven rearrangement of biological cells},
  Author                   = {James A. Glazier and Franqois Graner},
  Journal                  = {Phys. Rev. E},
  Year                     = {1993},
  Number                   = {3},
  Pages                    = {2128--2154},
  Volume                   = {47}
}

@Article{Graner:1992,
  Title                    = {Simulation of Biological Cell Sorting Using a Two-Dimensional Extended Potts Model},
  Author                   = {Fran\cois Graner and James A. Glazier},
  Journal                  = {Phys.~Rev.~Lett.},
  Year                     = {1992},
  Number                   = {13},
  Pages                    = {2013--2016},
  Volume                   = {69}
}

@Article{Hanke:2013,
  Title                    = {Understanding collective dynamics of soft active colloids by binary scattering},
  Author                   = {Timo Hanke and Christoph A. Weber and Erwin Frey},
  Journal                  = {Phys. Rev. E},
  Year                     = {2013},
  Pages                    = {052309},
  Volume                   = {88}
}

@Article{Hashimura:2019a,
  Title                    = {Collective cell migration of Dictyostelium without cAMP oscillations at multicellular stages},
  Author                   = {Hidenori Hashimura and Yusuke V. Morimoto and Masato Yasui and Masahiro Ueda},
  Journal                  = {Communications Biology},
  Year                     = {2019},
  Pages                    = {34},
  Volume                   = {2}
}

@Article{Hiraoka:2017,
  author    = {Takayuki Hiraoka and Takashi Shimada and Nobuyasu Ito},
  journal   = {J. Phys: Conf. Series},
  title     = {Collective motion in repulsive self-propelled particles in confined geometries},
  year      = {2017},
  pages     = {012006},
  volume    = {921},
}

@Article{Hiraoka:2016,
  author    = {T. Hiraoka and T Shimada and N Ito},
  journal   = {Phys. Rev. E},
  title     = {Order-disorder transition in repulsive self-propelled particle systems},
  year      = {2016},
  pages     = {062612},
  volume    = {94},
}

@Article{Hirashima:2017,
  author  = {T.~Hirashima and E.~G.~Rens and R.~M.~H. Merks},
  journal = {Dev. Growth Differ.},
  title   = {Cellular Potts modeling of complex multicellular behaviors in tissue morphogenesis},
  year    = {2017},
  number  = {5},
  pages   = {329--339},
  volume  = {59},
}

@Article{Kopf:2015,
  Title                    = {Collective cell migration by mechanical stress and substrate adhesiveness},
  Author                   = {Michael H. K\"{o}pf},
  Journal                  = {Phys.~Rev.~E},
  Year                     = {2015},
  Number                   = {1},
  Pages                    = {012712},
  Volume                   = {91}
}

@Article{Kopf:2013,
  Title                    = {A continuum model of epithelial spreading},
  Author                   = {Michael H. K\"{o}pf and Len M. Pismen},
  Journal                  = {Soft Matter},
  Year                     = {2013},
  Pages                    = {3727},
  Volume                   = {9}
}

@Article{Kabla:2012,
  Title                    = {Collective cell migration: leadership, invasion and segregation},
  Author                   = {Alexandre J. Kabla},
  Journal                  = {J. R. Soc. Interface},
  Year                     = {2012},
  Number                   = {77},
  Pages                    = {3268--3278},
  Volume                   = {9}
}

@Article{kuwayama:2013,
  Title                    = {Biological soliton in multicellular movement},
  Author                   = {Hidekazu Kuwayama and Shuji Ishida},
  Journal                  = {Sci. Rep.},
  Year                     = {2013},
  Pages                    = {2272},
  Volume                   = {3}
}

@Article{Lober:2015,
  Title                    = {Collisions of deformable cells lead to collective migration},
  Author                   = {Jakob L\"{o}ber and Falko Ziebert and Igor S. Aranson},
  Journal                  = {Sci. Rep.},
  Year                     = {2015},
  Pages                    = {9172},
  Volume                   = {5}
}

@Article{Leoni:2017,
  Title                    = {Model of Cell Crawling Controlled by Mechanosensitive Adhesion},
  Author                   = {M. Leoni and P. Sens},
  Journal                  = {Phys. Rev. Lett.},
  Year                     = {2017},
  Pages                    = {228101},
  Volume                   = {118}
}

@Article{Li:2008,
  Title                    = {Persistent Cell Motion in the Absence of External Signals: A Search Strategy for Eukaryotic Cells},
  Author                   = {L. Li and S. F. N\mbox{\o}rrelykke and E. C. Cox},
  Journal                  = {PLoS One},
  Year                     = {2008},
  Number                   = {5},
  Pages                    = {e2093},
  Volume                   = {3}
}

@Article{Lois:1994,
  Title                    = {Long-distance neuronal migration in the adult mammalian brain},
  Author                   = {C. Lois and A. Alvarez-Buylla},
  Journal                  = {Science},
  Year                     = {1994},
  Number                   = {5162},
  Pages                    = {1145--1148},
  Volume                   = {264}
}

@Article{Muller:1978,
  Title                    = {A specific glycoprotin as the target site of adhesion blocking Fab in aggregating Dictyostelium cells},
  Author                   = {Kurt M\"{u}ller and G\"{u}nter Gerisch},
  Journal                  = {Nature (London)},
  Year                     = {1978},
  Pages                    = {445--449},
  Volume                   = {274}
}

@Article{Maeda:2008,
  Title                    = {Ordered Patterns of Cell Shape and Orientational Correlation during Spontaneous Cell Migration.},
  Author                   = {Y. T. Maeda and J. Inose and M. Y. Matsuo and S. Iwaya and M. Sano},
  Journal                  = {PLoS ONE},
  Year                     = {2008},
  Number                   = {11},
  Pages                    = {e3734},
  Volume                   = {3}
}

@Article{Marchetti:2013,
  Title                    = {Hydrodynamics of soft active matter},
  Author                   = {M. C. Marchetti and J. F. Joanny and S. Ramaswamy and T. B. Liverpool and J. Prost and Madan Rao and R. Aditi Simha},
  Journal                  = {Rev. Mod. Phys.},
  Year                     = {2013},
  Pages                    = {1143},
  Volume                   = {85}
}

@Article{Matsushita:2020a,
  author    = {Katsuyoshi Matsushita},
  journal   = {Phys. Rev. E},
  title     = {Adhesion-inducing Long-Distance Transport of Cells on Tissue Surface},
  year      = {2020},
  month     = {May},
  number    = {5},
  pages     = {052410},
  volume    = {101},
}

@Article{Matsushita:2018,
  Title                    = {Emergence of collective propulsion through cell-cell adhesion},
  Author                   = {K. Matsushita},
  Journal                  = {Phys. Rev. E},
  Year                     = {2018},
  Number                   = {4},
  Pages                    = {042413},
  Volume                   = {97}
}

@Article{Matsushita:2017,
  author    = {K. Matsushita},
  journal   = {Phys. Rev. E},
  title     = {Cell-alignment patterns in the collective migration of cells with polarized adhesion},
  year      = {2017},
  pages     = {032415},
  volume    = {95},
}

@Article{Matsushita:2019,
  Title                    = {Cell Motion Alignment as Polarity Memory Effect},
  Author                   = {Katsuyoshi Matsushita and Kazuya Horibe and Naoya Kamamoto and Koichi Fujimoto},
  Journal                  = {J. Phys. Soc. Jpn.},
  Year                     = {2019},
  Pages                    = {103801},
  Volume                   = {88}
}

@Article{Merks:2005,
  Title                    = {A cell-centered approach to developmental biology},
  Author                   = {Roeland M.H. Merks and James A. Glazier},
  Journal                  = {Physica A},
  Year                     = {2005},
  Pages                    = {113--130},
  Volume                   = {352}
}

@Article{Najem:2016,
  Title                    = {Phase-field model for collective cell migration},
  Author                   = {Sara Najem and Martin Grant},
  Journal                  = {Phys. Rev. E},
  Year                     = {2016},

  Month                    = {May},
  Pages                    = {052405},
  Volume                   = {93}
}

@Article{Nakajima:2011,
  Title                    = {Kinetics of the cellular Potts model revisited},
  Author                   = {Akihiko Nakajima and Shuji Ishihara},
  Journal                  = {New J. Phys.},
  Year                     = {2011},
  Pages                    = {033035},
  Volume                   = {13}
}

@Article{Ohta:2017,
  author    = {Takao Ohta},
  journal   = {J. Phys. Soc. Jpn.},
  title     = {Dynamics of Deformable Active Particles},
  year      = {2017},
  number    = {10},
  pages     = {1856--1867},
  volume    = {86},
}

@Article{Peruani:2006,
  Title                    = {Nonequilibrium clustering of self-propelled rods},
  Author                   = {Fernando Peruani and Andreas Deutsch and Markus B{\"a}r},
  Journal                  = {Phys. Rev. E},
  Year                     = {2006},
  Pages                    = {030904(R)},
  Volume                   = {74}
}

@Article{Rappel:1999,
  Title                    = {Self-organized Vortex State in Two-Dimensional Dictyostelium Dynamics},
  Author                   = {Wouter-Jan Rappel and Alastair Nicol and Armand Sarkissian and Herbert Levine and William F. Loomis},
  Journal                  = {Phys. Rev. Lett.},
  Year                     = {1999},
  Pages                    = {1247},
  Volume                   = {83}
}

@Article{Sato:2015a,
  Title                    = {Cell Chirality Induces Collective Cell Migration in Epithelial Sheets},
  Author                   = {Katsuhiko Sato and Tetsuya Hiraiwa and Tatsuo Shibata},
  Journal                  = {Phys. Rev. Lett.},
  Year                     = {2015},
  Pages                    = {188102},
  Volume                   = {115}
}

@Article{Safran:2013,
  Title                    = {Physics of adherent cells},
  Author                   = {Ulrich S. Schwarz and Samuel A. Safran},
  Journal                  = {Rev. Mod. Phys.},
  Year                     = {2013},
  Pages                    = {1327--1372},
  Volume                   = {85}
}

@Book{Scianna:2013,
  Title                    = {Cellular Potts Model},
  Author                   = {Marco Scianna and Luigi Preziosi},
  Publisher                = {CRC Press, UK},
  Year                     = {2013}
}

@Article{Szabo:2006,
  Title                    = {Phase Transition in collective migration of tissue cells: experiment and model},
  Author                   = {B. Szab\'{o} and G. J. Szollosi and B. Gonci and Zs. Juranyi and D. Selmeczi and Tamas Vicsek},
  Journal                  = {Phys.~Rev.~E},
  Year                     = {2006},
  Pages                    = {061908},
  Volume                   = {74}
}

@Article{Takagi:2008,
  author  = {Hiroaki Takagi and Masayuki J. Sato and Toshio Yanagida and Masahiro Ueda},
  journal = {PLoS One},
  title   = {Functional Analysis of Spontaneous Cell Movement},
  year    = {2008},
  pages   = {e2648},
  volume  = {3},
}

@Article{Takeichi:2014,
  Title                    = {Dynamic contacts: rearranging adherens junctions to drive epithelial remodelling},
  Author                   = {Masatoshi Takeichi},
  Journal                  = {Nat. Rev. Mol. Cell. Biol.},
  Year                     = {2014},
  Pages                    = {397--410},
  Volume                   = {15}
}

@Article{Vicsek:1995,
  Title                    = {Novel Type of Phase Transition in a System of Self-Driven Particles},
  Author                   = {Tam\'{a}s Vicsek and Andr\'{a}s Czir\'{o}k and Eshel Ben-Jacob and Inon Cohen and Ofer Shochet},
  Journal                  = {Phys. Rev. Lett.},
  Year                     = {1995},
  Number                   = {6},
  Pages                    = {1226--1229.},
  Volume                   = {75}
}

@Article{Vroomans:2015,
  Title                    = {Segment-Specific Adhesion as a Driver of Convergent Extension},
  Author                   = {Renske M. A. Vroomans and Paulien Hogeweg and Kirsten H. W. J. ten Tusscher},
  Journal                  = {PLoS Comput. Biol.},
  Year                     = {2015},
  Number                   = {2},
  Pages                    = {e1004092},
  Volume                   = {11}
}

@Article{Wang:2000,
  Title                    = {The membrane glycoprotein gp150 is encoded by the lagC gene and mediates cellﾃδεつεδづつεδεつづδづつεδεつεδづつづδεつづδづつ｢ﾃ�?????ﾃδεつεδづつεδεつづδづつづδεつεδづつづδεつづδづつ田ell adhesion by heterophilic binding during Dictyostelium development.},
  Author                   = {J. Wang and L. Hou and D. Awrey and W. F. Loomis and R. A. Firtel and C.-H. Siu},
  Journal                  = {Dev. Biol.},
  Year                     = {2000},
  Pages                    = {734--745},
  Volume                   = {227}
}

@Article{Weber:2013,
  Title                    = {Long-Range Ordering of Vibrated Polar Disks},
  Author                   = {C. A. Weber and T. Hanke and J. Deseigne and S. L\'{e}onard and O. Dauchot and E. Frey and H. Chat\'{e}},
  Journal                  = {Phys. Rev. Lett.},
  Year                     = {2013},
  Pages                    = {208001},
  Volume                   = {110}
}

@Article{Weijer:2009,
  author  = {Cornelis J. Weijer},
  journal = {J. Cell Sci.},
  title   = {Collective cell migration in development},
  year    = {2009},
  pages   = {3215--3223},
  volume  = {122},
}

@Article{Yabunaka:2017b,
  Title                    = {Emergence of epithelial cell density waves},
  Author                   = {Shunsuke Yabunaka and Philippe Marcq},
  Journal                  = {Soft Matter},
  Year                     = {2017},
  Pages                    = {7046--7052},
  Volume                   = {13}
}

@Article{Zajac:2003,
  author  = {Mark Zajac and Gerald L. Jonesa and James A. Glazier},
  journal = {J. Theor. Biol.},
  title   = {Simulating convergent extension by way of anisotropic differential adhesion},
  year    = {2003},
  number  = {2},
  pages   = {247--259},
  volume  = {222},
}

@Article{Hakim:2017,
  author  = {V. Hakim and P. Silberzan},
  journal = {Rep. Prog. Phys.},
  title   = {Collective cell migration: a physics perspective},
  year    = {2017},
  pages   = {076601},
  volume  = {80},
}

@Article{Alert:2020,
  author  = {R. Alert and X. Trepat},
  journal = {Annu. Rev. Condens. Matter Phys.},
  title   = {Physical Models of Collective Cell Migration},
  year    = {2020},
  pages   = {77--101},
  volume  = {11},
}

@Article{Hiraiwa:2020,
  author  = {T. Hiraiwa},
  journal = {Phys. Rev. Lett.},
  title   = {Dynamic Self-Organization of Idealized Migrating Cells by Contact Communication},
  year    = {2020},
  pages   = {268104},
  volume  = {125},
}

@Article{Pascalis:2017,
  author  = {Chiara De Pascalis and Sandrine Etienne-Manneville},
  journal = {Mol. Biol. Cell},
  title   = {Single and collective cell migration: the mechanics of adhesions},
  year    = {2017},
  number  = {14},
  pages   = {1833},
  volume  = {28},
}

@Article{Matsushita:2021a,
  author  = {K. Matsushita and S. Yabunaka and K. Fujimoto},
  journal = {J. Phys. Soc. Jpn.},
  title   = {Polarity Fluctuation Inhibition by Memory in Collective Cell Motion},
  year    = {2021},
  number  = {5},
  pages   = {054801},
  volume  = {90},
}

@Article{Matsushita:2022a,
  author  = {K. Matsushita and H. Hashimura and H. Kuwayama and K. Fujimoto},
  journal = {J. Phys. Soc. Jpn.},
  title   = {Collective Cell Movement under Cell-Scale Tension Gradient at Tissue Interface},
  year    = {2022},
  number  = {5},
  pages   = {054802},
  volume  = {91},
}

@Article{Hayakawa:2020,
  author  = {Masayuki Hayakawa and Tetsuya Hiraiwa and Yuko Wada and Hidekazu Kuwayama and Tatsuo Shibata},
  journal = {eLife},
  title   = {Polar pattern formation induced by contact following locomotion in a multicellular system},
  year    = {2020},
  pages   = {e53609},
  volume  = {9},
}

@Article{Merks:2008a,
  author  = {Roeland M. H. Merks and Sergey V. Brodsky and Michael S. Goligorksy and Stuart A. Newman and James A. Glazier},
  journal = {Developmental Biology},
  title   = {Cell elongation is key to in silico replication of in vitro vasculogenesisand subsequent remodeling},
  year    = {2008},
  pages   = {44--54},
  volume  = {289},
}

@Article{Trepat:2009,
  author  = {Xavier Trepat and Michael R. Wasserman and Thomas E. Angelini and Emil Millet and David A. Weitz and James P. Butler and Jeffrey J. Fredberg},
  journal = {Nat. Phys.},
  title   = {Physical forces during collective cell migration},
  year    = {2009},
  pages   = {426--430},
  volume  = {5},
}

@Article{Haga:2005,
  author  = {Hisashi Haga and Chikako Irahara and Ryo Kobayashi and Toshiyuki Nakagaki and Kazushige Kawabata},
  journal = {biophys. J.},
  title   = {Collective Movement of Epithelial Cells on a Collagen Gel Substrate},
  year    = {2005},
  number  = {3},
  pages   = {2250--2256},
  volume  = {88},
}

@Article{Bi:2016,
  author  = {Dapeng Bi and Xingbo Yang and M. Cristina Marchetti and M. Lisa Manning},
  journal = {Phys. Rev. X},
  title   = {Motility-Driven Glass and Jamming Transitions in Biological Tissues},
  year    = {2016},
  pages   = {021011},
  volume  = {6},
}

@Article{Matsumoto:2024,
  author  = {Mami Matsumoto and Katsuyoshi Matsushita and Masaya Hane and Chentao Wen and Chihiro Kurematsu and Haruko Ota and Huy Bang Nguyen and Truc Quynh Thai and Vicente Herranz-P\'{e}rez and Masato Sawada and Koichi Fujimoto and Jos\'{e} Manuel García-Verdugo and Koutarou D Kimura and Tatsunori Seki and Chihiro Sato and Nobuhiko Ohno and Kazunobu Sawamoto},
  journal = {EMBO Mol. Med.},
  title   = {Neuraminidase inhibition promotes the collective migration of neurons and recovery of brain function},
  year    = {2024},
  month   = jun,
  number  = {6},
  pages   = {1228--1253},
  volume  = {16},
}

@Article{Sawada:2013,
  author  = {Masato Sawada and Kazunobu Sawamoto},
  journal = {The Keio Journal of Medicine},
  title   = {Mechanisms of Neurogenesis in the Normal and Injured Adult Brain},
  year    = {2013},
  number  = {1},
  pages   = {13--28},
  volume  = {62},
}

@Article{Fujioka:2017,
  author  = {T.~Fujioka and N.~Kaneko and I.~Ajioka and K.~Nakaguchi and T.~Omata and H.~Ohba and R.~F\"assler and J.~M.~Garc\'ia-Verdugo and K.~Sekiguchi and N.~Matsukawa and K.~Sawamoto},
  journal = {EBioMedicine},
  title   = {\beta1 integrin signaling promotes neuronal migration along vascular scaffolds in the post-stroke brain.},
  year    = {2017},
  pages   = {195--203},
  volume  = {16},
}

@Article{Johnson:2005,
  author  = {C.~P.~Johnson and I.~Fujimoto, U.~Rutishauser, D.~E.~Leckband},
  journal = {J. Biol. Chem.},
  title   = {Direct evidence thatneural cell adhesion molecule (NCAM) polysialylation increasesintermembrane repulsion and abrogates adhesion.},
  year    = {2005},
  pages   = {137--145},
  volume  = {280},
}

@Article{Ajioka:2015,
  author  = {Itsuki Ajioka and Hideo Jinnou and Kei Okada and Masato Sawada and Shinji Saitoh and Kazunobu Sawamoto},
  journal = {Tissue Engineering Part A},
  title   = {Enhancement of Neuroblast Migration into the Injured Cerebral Cortex Using Laminin-Containing Porous Sponge},
  year    = {2015},
  number  = {1-2},
  pages   = {193--201},
  volume  = {21},
}

@Article{Matsushita:2024,
  author  = {Katsuyoshi Matsushita and Taiko Arakaki and Koichi Fujimoto},
  journal = {J. Phys. Soc. Jpn.},
  title   = {Motion Ordering in Cellular Polar–Polar and Polar–Nonpolar Interactions},
  year    = {2024},
  pages   = {114801},
  volume  = {93},
}

@Article{Kaneko:2018,
  author  = {N.~Kaneko and V.~Herranz-P\'erez and T.~Otsuka and H.~Sano and N.~Ohno and T.~Omata and H.~B.~Nguyen and T.~Q.~Thai and A.~Nambu and Y.~Kawaguchi and J.~M.~Garc\'ia-Verdugo, K.~Sawamoto},
  journal = {Sci. Adv.},
  title   = {New neurons use Slit-Robo signaling to migrate through the glial meshwork and approach a lesion for functional regeneration},
  year    = {2018},
  number  = {12},
  pages   = {eaav0618},
  volume  = {4},
}

@Article{Mogilner:2012,
  author  = {Alex Mogilner and Jun Allard and Roy Wollman},
  journal = {Science},
  title   = {Cell Polarity: Quantitative Modelingas a Tool in Cell Biology},
  year    = {2012},
  pages   = {175--179},
  volume  = {336},
}

@Article{Bertrand:2024,
  author  = {Thibault Bertrand and Joseph d'Alessandro and Ananyo Maitra and Shreyansh Jain and Barbara Mercier and Ren\'e-Marc M\`ege and Benoit Ladoux and Rapha\"el Voituriez},
  journal = {Phys. Rev. Res.},
  title   = {Clustering and ordering in cell assemblies with generic asymmetric aligning interactions},
  year    = {2024},
  pages   = {023022},
  volume  = {6},
}

@Article{McDonald:2003,
  author  = {J.~A.~McDonald and E.~M.~Pinheiro and D.~J.~Montell},
  journal = {Development},
  title   = {PVF1, a PDGF/VEGF homolog, is sufficient to guide border cells and interacts genetically with taiman},
  year    = {2003},
  number  = {15},
  pages   = {3469--3478},
  volume  = {130},
}

@Article{Pfister:2016,
  author  = {Katherine Pfister and David R. Shook and Chenbei Chang and Ray Keller and Paul Skoglund},
  journal = {Development},
  title   = {Molecular model for force production and transmission during vertebrate gastrulation},
  year    = {2016},
  number  = {4},
  pages   = {715–727},
  volume  = {143},
}

@Article{Belmonte:2016,
  author  = {Julio M. Belmonte and Maciej H. Swat and James A. Glazier},
  journal = {PLS Comput. Biol.},
  title   = {Filopodial-Tension Model of ConvergentExtension of Tissues},
  year    = {2016},
  number  = {6},
  pages   = {e1004952},
  volume  = {12},
}

@Book{Alberts:2022,
  author    = {Bruce Alberts and Rebecca Heald and Alexander Johnson and David Morgan and Martin Raff},
  publisher = {W. W. Norton \& Company},
  title     = {Molecular Biology of the Cell 7th ed.},
  year      = {2022},
}

@Article{Khalila:2010,
  author  = {Antoine A. Khalila and Peter Friedl},
  journal = {Integr. Biol.},
  title   = {Determinants of leader cells in collective cell migration},
  year    = {2010},
  pages   = {568–-574},
  volume  = {2},
}

@Article{Belvindrah:2007,
  author  = {Richard Belvindrah and Sabine Hankel and John Walker and Bruce L. Patton and and Ulrich M\"uller},
  journal = {The Journal of Neuroscience},
  title   = {{\beta}1-Integrins Control the Formation of Cell Chains in theAdult Rostral Migratory Stream},
  year    = {2007},
  number  = {10},
  pages   = {2704--2717},
  volume  = {27},
}

@Article{Fujiike:2018,
  author  = {Kazuma Fujikake and Masato Sawada and Takao Hikita and Yayoi Seto and Naoko Kaneko and Vicente Herranz-Pérez and Natsuki Dohi and Natsumi Homma and, Satoshi Osaga and Yuchio Yanagawa and Toshihiro Akaike and Jose Manuel García-Verdugo and Mitsuharu Hattori and Kazuya Sobue and Kazunobu Sawamoto},
  journal = {J Neurosci.},
  title   = {Detachment of Chain-Forming Neuroblasts by Fyn-Mediated Control of cell-cell Adhesion in the Postnatal Brain},
  year    = {2018},
  number  = {19},
  pages   = {4598--4609},
  volume  = {38},
}

@Article{Reffey:2014,
  author  = {M. Reffay and M. C. Parrini and O. Cochet-Escartin and B. Ladoux and A. Buguin and S. Coscoy and F. Amblard and J. Camonis and P. Silberzan},
  journal = {Natl. Biol. Cell},
  title   = {Interplay of RhoA and mechanical forces in collectivecell migration driven by leader cells},
  year    = {2014},
  number  = {3},
  pages   = {217--223},
  volume  = {16},
}

@Article{Khataee:2021,
  author  = {Hamid Khataee and Andras Czirok and Zoltan Neufeld},
  journal = {Phys. Rev. E},
  title   = {Contact inhibition of locomotion generates collective cell migration without chemoattractants in an open domain},
  year    = {2021},
  pages   = {014405},
  volume  = {104},
}

@Article{Schneyder:2017,
  author  = {Simon K. Schnyder and John J. Molina and Yuki Tanaka and Ryoichi Yamamoto},
  journal = {Sci Rep.},
  title   = {Collective motion of cells crawling on a substrate: roles of cell shape and contact inhibition},
  year    = {2017},
  pages   = {5163},
  volume  = {7},
}

@Article{Camley:2017,
  author  = {B. A. Camley and W.-J. Rappel},
  journal = {Journal of Physics D: Applied Physics},
  title   = {Physical models of collective cell motility: from cell to tissue},
  year    = {2017},
  pages   = {113002},
  volume  = {50},
}

@Article{Luster:2005,
  author  = {Andrew D Luster and Ronen Alon and Ulrich H von Andrian},
  journal = {Nature Immunology},
  title   = {Immune cell migration in inflammation: present and future therapeutic targets},
  year    = {2005},
  pages   = {1182–1190},
  volume  = {6,},
}

@Article{Novikova:2017,
  author  = {Elizaveta A. Novikova and Matthew Raab and Dennis E. Discher and Cornelis Storm},
  journal = {Phys. Rev. Lett.},
  title   = {Persistence-Driven Durotaxis: Generic, Directed Motility in Rigidity Gradients},
  year    = {2017},
  pages   = {078103},
  volume  = {118},
}

@Article{Varennes:2017,
  author  = {Julien Varennes and Sean Fancher and Bumsoo Han and Andrew Mugler},
  journal = {Phys. Rev. Lett.},
  title   = {Emergent versus Individual-Based Multicellular Chemotaxis},
  year    = {2017},
  number  = {18},
  pages   = {188101},
  volume  = {119},
}

@Article{Marzio:2025,
  author  = {Margherita De Marzio and Amit Das and Jeffrey J. Fredberg and Dapeng Bi},
  journal = {Phys. Rev. Lett.},
  title   = {Epithelial Layer Fluidization by Curvature-Induced Unjamming},
  year    = {2025},
  pages   = {138402},
  volume  = {134},
}

@Article{Sadhukan:2025,
  author  = {Shubhadeep Sadhukhan and Cristina Martinez-Torres and Samo Peni\v{c} and Carsten Beta and Ale\v{s} Igli\v{c} and Nir Gov},
  journal = {Phys. Rev. Lett.},
  title   = {Modeling how lamellipodia-driven cells maintain persistent migration and interactwith external barriers},
  year    = {2025},
  pages   = {013319},
  volume  = {7},
}

@Article{Wang:2025,
  author  = {Wei Wang and Robert A. Law and Emiliano Perez Ipi\~na Konstantinos Konstantopoulos and Brian A. Camley},
  journal = {Phys. Rev. X},
  title   = {Confinement, Jamming, and Adhesion in Cancer Cells Dissociating from a Collectively Invading Strand},
  year    = {2025},
  pages   = {013012},
  volume  = {3},
}

@Article{Pinto:2022,
  author  = {Diogo E. P. Pinto and Margarida M. Telo da Gama and Nuno A. M. Ara\'ujo},
  journal = {Phys. Rev. Res.},
  title   = {Cell motility in confluent tissues induced by substrate disorder},
  year    = {2022},
  pages   = {023186},
  volume  = {4},
}

@Article{fujimori:2024,
  author  = {Taihei Fujimori and Hidenori Hashimura and Satoshi Sawai},
  journal = {Methods Mol Biol.},
  title   = {Imaging-Based Analysis of Cell-Cell Contact-Dependent Migration in Dictyostelium},
  year    = {2024},
  pages   = {23--36},
  volume  = {2828},
}

@Article{Hayakawa:2024,
  author  = {M.~Hayakawa and H.~Kuwayama and T.~Shibata},
  journal = {Methods Mol Biol.},
  title   = {A Mutant of Dictyostelium discoideum, KI-Cell, as a Model of Collective Cell Migration Independent of Chemotaxis.},
  year    = {2024},
  pages   = {37--43},
  volume  = {2828},
}

@Article{Merks:2008b,
  author  = {Roeland M. H. Merks and Erica D. Perryn and Abbas Shirinifard and James A. Glazier},
  journal = {Plos Comput. Biol.},
  title   = {Contact-inhibited chemotaxis in de novo and sprouting blood-vessel growth},
  year    = {2008},
  number  = {9},
  pages   = {e1000163},
  volume  = {4},
}

@Article{Palm:2013,
  author  = {Margriet M. Palm and Roeland M. H. Merks},
  journal = {Phys. Rev. E},
  title   = {Vascular networks due to dynamically arrested crystalline ordering of elongated cells},
  year    = {2013},
  pages   = {012725},
  volume  = {87},
}

@Article{Kafer:2007,
  author  = {Jos K\"afer and Takashi Hayashi and Athanasius F. M. Mar\'ee and Fran\c{c}ois Graner},
  journal = {Proc. Natl. Acad. Sci. USA},
  title   = {Cell adhesion and cortex contractility determinecell patterning in the Drosophila retina},
  year    = {2007},
  number  = {47},
  pages   = {18549--18554},
  volume  = {104},
}

@Article{Savill:1997,
  author  = {Nicholas J. Savill and Paulien Hogeweg},
  journal = {J. Theor. Biol.},
  title   = {Modelling Morphogenesis: From Single Cells to Crawling Slugs},
  year    = {1997},
  number  = {3},
  pages   = {229--235},
  volume  = {184},
}

@Article{Grazier:2008,
  author  = {James A. Glazier and Ying Zhang and Maciej Swat and Benjamin Zaitlen and Santiago Schnell},
  journal = {Curr. Topics in Develop. Biol.},
  title   = {Coordinated Action of N-CAM, N-cadherin, EphA4, and ephrinB2 Translates Genetic Prepatterns into Structure during Somitogenesis in Chick},
  year    = {2008},
  pages   = {205--247},
  volume  = {81},
}

@Article{Graizer:2011,
  author  = {Susan D. Hester and Julio M. Belmonte and J. Scott Gens and Sherry G. Clendenon and James A. Glazier},
  journal = {PLoS Comput. Biol.},
  title   = {A Multi-cell, Multi-scale Model of Vertebrate Segmentation and Somite Formation},
  year    = {2011},
  number  = {10},
  pages   = {e1002155},
  volume  = {7},
}

@Article{Belousov:2024,
  author  = {R. Belousov and S. Savino and P. Moghe and T. Hiiragi and L. Rondoni and A. Erzberger},
  journal = {Phys. Rev. Lett.},
  title   = {Poissonian Cellular Potts Models Reveal Nonequilibrium Kinetics of Cell Sorting},
  year    = {2024},
  pages   = {248401},
  volume  = {132},
}

@Article{Braat:2024,
  author  = {Quirine J. S. Braat and Cornelis Storm and Liesbeth M. C. Janssen},
  journal = {Phys. Rev. E},
  title   = {Formation of motile cell clusters in heterogeneous model tumors: The role of cell-cell alignment},
  year    = {2024},
  pages   = {064401},
  volume  = {110},
}

@Article{Wurthner:2026,
  author  = {Laeschkir W\"urthner and Frederik Graw},
  journal = {Phys. Rev E},
  title   = {Geometry of disordered porous environments regulates cell migration},
  year    = {2026},
  pages   = {014407},
  volume  = {113},
}

@Article{Noureen:2025,
  author  = {Shahzeb Raja Noureen and Richard L. Mort and Christian A. Yates},
  journal = {Phys. Rev. E},
  title   = {Modeling adhesion in stochastic and mean-field models of cell migration},
  year    = {2025},
  pages   = {014419},
  volume  = {111},
}

@Article{Siu:2011,
  author  = {Chi-Hung Siu and Shrivani Sriskanthadevan and Jun Wang and Liansheng Hou and Gong Chen and Xiaoqun Xu and Alexander Thomson and Chunxia Yang},
  journal = {Dev. Growth Differ.},
  title   = {Regulation of spatiotemporal expression of cell–celladhesion molecules during development of Dictyosteliumdiscoideum},
  year    = {2011},
  pages   = {518–-527},
  volume  = {53},
}

@Article{Saito:2024,
  author  = {Nen Saito and Shuji Ishihara},
  journal = {Sci. Adv.},
  title   = {Cell deformability drives fluid-to-fluid phase transition in active cell monolayers},
  year    = {2024},
  pages   = {eadi8433},
  volume  = {10},
}

\end{document}